\documentclass[final,2p,times,twocolumn]{elsarticle}

\usepackage{amssymb}

\begin{document}

\begin{frontmatter}



\title{Superstatistical cluster decay}


\author[NCBJ]{Grzegorz Wilk}
\ead{wilk@fuw.edu.pl}


\address[NCBJ]{National Centre for Nuclear Research,
        Department of Fundamental Research, Ho\.za 69, 00-681
        Warsaw, Poland}

\author[JKU]{Zbigniew W\l odarczyk}
\ead{zbigniew.wlodarczyk@ujk.edu.pl}

\address[JKU]{Institute of Physics, Jan Kochanowski University,
\'Swi\c{e}tokrzyska 15, 25-406 Kielce, Poland}

\begin{abstract}
We provide an overview of Tsallis statistics, presented as a
special case of superstatistics and applied to the multiparticle
processes described by the statistical cluster model. This model
combines Boltzman statistics applied to hadronization of clusters
and superstatistics induced by fluctuations of their Lorentz
factors. It results in a Tsallis-like distribution of transverse
momenta with some scale, the origin of which is discussed. We show
that this distribution becomes a Tsallis distribution in a special
case, namely when one combines fluctuations of the Lorentz factor
and temperature inside the cluster, given by beta and gamma
distributions, respectively.

\end{abstract}


\end{frontmatter}

\section{Introduction}

Quasi-power like distributions, exemplified by the Tsallis
distribution \cite{Tsallis},
\begin{equation}
h(E) = \frac{2-q}{T}\left[ 1 -
(1-q)\frac{E}{T}\right]^{\frac{1}{1-q}} \label{Tsallis}
\end{equation}
are nowadays recognized as a legitimate tool for use in describing
different distributions of this type (usually with $E$ replaced by
transverse momentum $p_T$). This is the more so when it was
realized that, albeit based on nonextensive statistical
considerations, it is equivalent to the so called Hagedorn
distribution \cite{H} (known in the field of particle
physics)\footnote{Actually, it was originally proposed in
\cite{M}.},
\begin{equation}
h(E) = \frac{n-1}{nT}\left( 1 + \frac{E}{nT}\right)^{-n},
\label{Hagedorn}
\end{equation}
where for $n = \frac{1}{q-1}$ both distribution coincide. The most
recent examples of different phenomenological applications of one
of these formulas can be found in \cite{RW,MCA,Biro,Indian,BYW}
(and references therein\footnote{One should mention that the first
papers applying $q$-statistics to scattering data in high energy
physics dates back to \cite{Bediaga} and \cite{CB}, the second
also pointing out the connection to Hagedorn's theory.}. It is
remarkable that such a distribution also fits recent LHC data on
transverse momentum distributions extending up to $p_T = 200$ GeV,
with cross sections dropping through $\sim 14$ orders of
magnitude, cf. for example \cite{LHC}. There are numerous ways
besides nonextensive statistics which lead to distributions
(\ref{Tsallis}) or (\ref{Hagedorn}), like, for example,
fluctuations of the scale parameter (especially the so called
superstatistics), preferential attachment or multiplicative noise
- to name only a few (see \cite{Zakopane,restWWrev} for recent
reviews)\footnote{Actually, the LHC results mentioned above were
also interpreted in the spirit of a Tsallis-like distribution,
emerging in this case, under some approximations, from the hard
collisions of partons described by QCD methods and followed by
showering/hadronization processes into hadrons \cite{qQCD}.}.

\section{Scale parameter fluctuations}

Let us concentrate on the case of superstatistics (understood as a
superposition of two different statistics relevant to driven
nonequilibrium systems with a stationary state and intensive
parameter fluctuations) \cite{SS}. It is based on the observation
that the Tsallis distribution (\ref{Tsallis}) appears in a quite
natural way as a result of $T'$ scale fluctuations in the BG
distribution,
\begin{equation}
f(E) = \frac{1}{T'}\exp\left( - \frac{E}{T'}\right), \label{BG}
\end{equation}
when these fluctuations are described by a gamma distribution in
the variable $(1/T')$ \cite{qWW,qBJ,CBGamma,restWWrev},
\begin{equation}
g\left(T'\right) = \frac{1}{nT\Gamma(n)}\left(
\frac{nT}{T'}\right)^{n+1} \exp\left( -\frac{nT}{T'}
\right),~\quad n=\frac{1}{q-1}, \label{Gd}
\end{equation}
where the parameter $q$ defines the strength of the $T'$
fluctuations,
\begin{equation}
 q = 1 + \frac{Var\left(1/T'\right)}{\langle 1/T'\rangle^2}. \label{SS}
 \end{equation}
It should be mentioned at this point that depending on the
statistical properties of the fluctuations, one obtains different
effective statistical mechanical descriptions. Tsallis statistics
follow from the above gamma distribution of an intensive variable,
but other classes of generalized statistics can be obtained as
well and, for small variance of the fluctuations, they all behave
in a universal way. From the thermal perspective, this corresponds
to a situation in which the system considered, for example a
decaying cluster, is not homogeneous, but has different
temperatures in different parts, which are fluctuating around some
mean temperature $T$ (an essentially identical result is obtained
when the temperature fluctuates from one cluster to another
remaining  constant in a given cluster). It must therefore be
described by two parameters: a mean temperature $T$ and the mean
strength of the fluctuations defined by $q$. The gamma
distribution (\ref{Gd}) itself emerges in a quite natural way,
either from thermal considerations corresponding to the purely
multiplicative noise \cite{qWW}, or in a stochastic approach
without correlation between the assumed additive and
multiplicative noises \cite{qBJ}\footnote{The first identification
of local temperature fluctuations as the source of Tsallis
distributions observed in high energy physics data was done in
\cite{qWW}, later on it was also used in analysis of data from
cosmic rays events in \cite{cosmicCB}.}.

We shall now discuss this situation in more detail. Note that the
distribution $g\left(T'\right)$ in Eq. \ref{Gd}) is in fact the
product of two distributions:
\begin{equation}
g\left(T'\right) \sim g_1\left(T'\right)\cdot g_2\left(T'\right),
\label{g1g2}
\end{equation}
where
\begin{equation}
g_1\left(T'\right) = \left( \frac{1}{T'}\right)^{\kappa}
\label{g1}
\end{equation}
is a scale free power law, and
\begin{equation}
g_2\left( T' \right) = \exp \left( - \frac{nT}{T'}\right),
\label{g2}
\end{equation}
is an exponential, which for small values of $T'$ cuts off the
distribution $g\left(T'\right)$, with the scale parameter $T$
determining how fast this cut-off is.

Suppose now that we fluctuate the BG distribution (\ref{BG}) using
only the $g_1\left(T'\right)$ part of the distribution
$g\left(T'\right)$. As a result we obtain the following scale free
distribution:
\begin{equation}
h_1(E) = \int_0^{\infty} dT'\, g_1\left(T'\right) \exp\left( -
\frac{E}{T'}\right) \propto E^{-\kappa + 1}. \label{h1}
\end{equation}
The scale appears when one cuts-off somehow the small values of
$T'$. For example, a sharp cut-off of the small $T'$, i.e.,
limiting $T'$ to $T' > T$ only, results in the following
distribution:
\begin{eqnarray}
h_2(E) &=& \int_T^{\infty} dT'\, g_1\left(T'\right) \exp\left( -
\frac{E}{T'}\right) \propto \nonumber\\
&\propto& E^{-\kappa + 1} \left[ \Gamma(\kappa - 1) - \Gamma
\left(\kappa - 1, \frac{E}{T}\right)\right], \label{h2}
\end{eqnarray}
where $\Gamma(x,y)$ is an incomplete gamma function. The factor
$\Gamma(\kappa - 1) - \Gamma(\kappa - 1, E/T)$ now suppresses the
power distribution (\ref{h1}) for small values of $E$. However,
the form of this suppression is not the same as in the Tsallis
distribution. This can be seen by comparing the expansion
\begin{eqnarray}
&&\!\!\!\!\!\!\!\!\!\!\!\!\!\! \left(
\frac{E}{T}\right)^{-\kappa}\cdot \left[ \Gamma(\kappa - 1) -
\Gamma \left(\kappa - 1, \frac{E}{T}\right)\right] =
\frac{1}{\kappa} +\nonumber\\
&&\qquad \quad  + \sum_{i=1}^{\infty}\frac{\Gamma(i +
\kappa)}{\Gamma(i + \kappa + 1)\Gamma(i + 1)}\left( -
\frac{E}{T}\right)^i \label{comp1}
\end{eqnarray}
with the corresponding expansion of the Tsallis factor:
\begin{equation}
\left( 1 + \frac{E}{\kappa T}\right)^{-\kappa} = \frac{1}{\kappa}
+  \sum_{i=1}^{\infty}\frac{\Gamma(i + \kappa)}{\Gamma(1 +
\kappa)\Gamma(i+1)}\left( - \frac{E}{\kappa T}\right)^i.
\label{comp2}
\end{equation}
However, if we smooth out this suppression by replacing the
previous sharp limitation of the integrand by some smooth
suppression factor, provided, for example, by the exponential
function $g_2\left( T'\right)$ from Eq. (\ref{g2}), we get, as
result, the Tsallis distribution defined in  Eq. (\ref{Tsallis}).

\section{Approximate description of cluster decay}

Recently, a new statistical cluster decay model of hadronization
has been analyzed numerically, showing that even without resorting
to approaches of the kind mentioned above, the resulting
distribution of transverse momenta follows rather closely a
Tsallis distribution \cite{AB}, albeit not identically so. In this
work the hadronic clusters produced were supposed to decay purely
thermally (i.e. by following the usual exponential Boltzmann-Gibbs
(BG) statistics) but, at the same time, were supposed to move in
the transverse direction with a fluctuating (transverse) Lorentz
factor distributed according to the assumed power law. It turns
out that the combination of both distributions follows (at least
numerically) a quasi-power like distribution, closely resembling a
Tsallis distribution. Note that, according to what was said before
regarding superstatistics, production and decay of such clusters
can be regarded as an example of superstatistics at work (and not
necessarily resulting in a Tsallis distribution).

In this work we shall investigate this phenomenon in more detail,
aiming to obtain its analytical justification and a deeper
understanding from the nonextensive statistical point of view. We
start with statistical cluster decay, discussed recently in
\cite{AB}. The distribution of transverse momenta proposed there
is\footnote{It is worth noting that cluster models have a very
long tradition, a similar formula was already derived in
\cite{OldC}.},
\begin{equation}
f(\left( p_T\right) \propto \int d\gamma_T K_0\left(\gamma_T
\frac{m_T}{T}\right)I_0\left(\sqrt{\gamma_T^2 -
1}\frac{p_T}{T}\right) g\left(\gamma_T\right), \label{AB1}
\end{equation}
where $m_T$ is the transverse mass and $\gamma_T$ denotes the
transverse Lorentz factor ($K_0$ and $I_0$ are modified Bessel
functions of the, respectively, second and first kind). This
factor is then supposed to fluctuate according to some power-like
distribution,
\begin{equation}
g\left(\gamma_T\right) \sim \gamma_T^{-\kappa},\label{gamTfluct}
\end{equation}
i.e. according to the $g_1$ function from Eq. (\ref{g1}) with $T'$
replaced by $\gamma_T$. Numerical calculations in \cite{AB} show
that whereas for large $p_T$ one has
\begin{equation}
f\left( p_T\right) \sim  p_T^{-\kappa - 1}, \label{AB2}
\end{equation}
there is a suppression for small values of $p_T$. To describe its
origin note that the $p_T$ distribution given by the product of
two Bessel functions from Eq. (\ref{AB1}),
\begin{equation}
\tilde{f}\left( p_T\right) \sim K_0\left(\gamma_T
\frac{m_T}{T}\right)I_0\left(\sqrt{\gamma_T^2 -
1}\frac{p_T}{T}\right), \label{tildef}
\end{equation}
depends strongly on the transverse Lorentz factor of the decaying
cluster, $\gamma_T$,
\begin{eqnarray}
\tilde{f}\left(p_T\right) &\sim& \left\{
\begin{array}{lll}
\exp\left (- p_T\right)&\quad {\rm for}\quad \gamma_T = 1,&\\
&&\\
\frac{1}{p_T} &\quad {\rm for~~large}~\gamma_T.&\\
\end{array}\right.
\label{AB3}
\end{eqnarray}
Because:
\begin{itemize}
\item $K_0(x)I_0(y) \simeq \frac{\exp(-x+y)}{2\sqrt{xy}}$ for $x,y
>> \frac{1}{4}$,
\item $\gamma_T - \sqrt{\gamma_T^2 - 1} \simeq
\frac{1}{2\gamma_T}$ for large $\gamma_T$\footnote{For $\gamma_T >
5$ the difference is less than $1$\%. We expand in series in
$z=1/\gamma_T$: $1/z-\sqrt{1/z^2-1}= 1/z - \sqrt{1 - z^2}/z \simeq
z/2$.}, \item $m_T \simeq p_T$ for $p_T >> m$,
\end{itemize}
therefore we can rewrite Eq. ({\ref{AB1}) in the following way:
\begin{eqnarray}
f\left( p_T\right) &\propto& \int_1^{\infty} d\gamma_T \exp\left(
- \frac{p_T}{T\gamma_T}\right) p_T^{-1} \gamma_T^{-\kappa - 1}
=\nonumber\\
&=& p_T^{-\kappa - 1}\left[\Gamma(\kappa) -
\Gamma\left(\kappa,\frac{p_T}{T}\right)\right].\label{AB4}
\end{eqnarray}
Note that because $\gamma_T$ is always limited to values
$\gamma_T\ge 1$ only, there is a natural cut-off in Eq.
(\ref{AB1}), which makes it similar and comparable to Eq.
(\ref{h2}) discussed above. As a result, here the power-like
distribution of $p_T$ is also suppressed in the region of small
values of transverse momenta. The factor $\left[\Gamma(\kappa) -
\Gamma \left(\kappa, p_T/T \right)\right]$ introduces the behavior
$\sim p_T^{\kappa}$ for small $p_T$ and remains constant for large
values of $p_T$. To summarize, such an approximation must
therefore lead to a result which is very near to the numerical
results presented in \cite{AB}, and not coinciding with a Tsallis
distribution.

Because of the important role played by fluctuations of the
Lorentz factor $\gamma_T$ described by the distribution
$g\left(\gamma_T\right)$ in Eq. (\ref{AB1}), it is interesting to
speculate on its possible origin emerging from the quark structure
of the colliding nucleons. Let us denote by $x=p_T/p_h$ the
fraction of transverse momentum of the parton, $p_T$, with respect
to the momentum of hadron, $p_h$, and let us assume that the
density of the parton distribution is
\begin{equation}
w(x) = Ax^a(1 - x)^b. \label{wx1c2}
\end{equation}
In the center of mass system the Lorentz factor of the cluster
formed by the collision of partons with fractions of momenta equal
to, respectively, $x_1$ and $x_2$, is given by
\begin{equation}
\gamma_T = \frac{x_1 + x_2}{2\sqrt{x_1 x_2}} .\label{LFofC}
\end{equation}
Notice that if $x_1 = x_2$ then $\gamma_T = 1$, it becomes greater
than unity only for nonsymmetric values of $x_{1,2}$. It is
convenient to change variables to
\begin{equation}
x_1 = \rho \cos \phi,\qquad x_2 = \rho\sin \phi \label{rhophi}
\end{equation}
in which
\begin{equation}
\frac{x_2}{x_1} = \tan \phi,\quad x_1^2 + x_2^2 = \rho^2\quad{\rm
and}\quad \gamma_T = \frac{1 + \tan \phi}{2\sqrt{\tan \phi}}.
\label{example}
\end{equation}
The Jacobian of this transformation is equal $|J| = \rho$. The
distribution of the Lorentz factor $\gamma_T$ is now given by
\begin{eqnarray}
\frac{1}{\sigma}\frac{d\sigma}{d\gamma_T}  &=& A^2 \int d\rho
\rho^{1 + 2a}\cdot \int d\phi \left( \frac{\tan \phi}{1 + \tan^2
\phi}
\right)^a \cdot\nonumber\\
&&\cdot\left(1 - \rho\frac{1 + \tan \phi}{\sqrt{1 + \tan^2 \phi}}
+ \rho^2 \frac{\tan \phi}{1 + \tan^2 \phi} \right)^b
\cdot\nonumber\\
&& \cdot \delta\left( \gamma_T - \frac{1 + \tan \phi}{2 \sqrt{\tan
\phi}}\right) \label{gTrphi}
\end{eqnarray}
Denoting
\begin{equation}
y = \gamma_T - \frac{1 + \tan \phi}{2 \sqrt{\tan \phi}}
\label{eqy}
\end{equation}
we can write
\begin{equation}
\int d\phi\, (\dots) \delta (y) = \int dy\, (\dots) \sum_i
\frac{\delta (y)}{\big | \frac{dy}{d\phi} \big|_{\phi_{0i}}}
\end{equation}
where $\phi_{0i}$ are solutions of Eq. (\ref{eqy}) for $y=0$. Now,
\begin{equation}
\frac{dy}{d\phi} = \frac{1 - \tan \phi}{4\tan \phi \sqrt{\tan
\phi}}\left( 1 + \tan^2 \phi \right)
\end{equation}
and there are two solutions, $\phi_{0(1,2)}$, for which we have
\begin{eqnarray}
&&\tan \phi_{0(1,2)} = 2\gamma_T^2 - 1 \pm 2\gamma_T
\sqrt{\gamma_T^2 - 1}, \label{tanphi0} \\
&&{\rm where}\quad \tan \phi_{01} \ge 1\quad {\rm and}\quad \tan
\phi_{02} \le 1. \label{GS}
\end{eqnarray}
It means therefore that Eq. (\ref{gTrphi}) has now two terms with
different limits of integrations over $\rho$ in each of them,
\begin{equation}
\rho \in \left(0, \rho_1\right);\quad \rho_1 = \frac{1}{\sin
\phi_{01}} = \frac{\sqrt{1 + \tan^2 \phi_{01}}}{\tan \phi_{01}}
\label{rho1}
\end{equation}
(corresponding to $\phi \in (\pi/4, \pi/2)$)
\begin{equation}
\rho \in \left(0, \rho_2\right);\quad \rho_2 = \frac{1}{\cos
\phi_{02}} = \sqrt{1 + \tan^2 \phi_{02}}. \label{rho2}
\end{equation}
(corresponding to $\phi \in (0, \pi/4)$). However, in both cases
we cover the whole range of $x_{1,2}$. For the choice (\ref{rho1})
it is done preserving all the time inequality $x_2 \ge x_1$,
whereas for the choice (\ref{rho2}) it is done by keeping all the
time $x_2 \leq x_1$. In both cases the final result must be the
same. Therefore, choosing for example, Eq. (\ref{rho2},) we have
that
\begin{eqnarray}
\!\!\!\!\!\!\!\frac{1}{\sigma}\frac{d\sigma}{d\gamma_T}\!\!\
&=&\!\!\ 2A^2 \frac{1}{\Big| \frac{dy}{d \phi} \Big|_{\phi_{02}}}
\left( \frac{\tan \phi_{02}}{1 + \tan^2 \phi_{02}} \right)^a
\int_0^{\rho_2} d\rho
\rho^{1 + 2a} \cdot \nonumber\\
\!\!\!\!\!\!\!\ &\cdot& \!\!\!\!\!\left(1 - \rho\frac{1 + \tan
\phi_{02}}{\sqrt{1 + \tan^2 \phi_{02}}} + \rho^2 \frac{\tan
\phi_{02}}{1 + \tan^2 \phi_{02}} \right)^b. \label{e1}
\end{eqnarray}
For $\gamma_T >> 1$ it can be numerically approximated by
\begin{equation}
\frac{1}{\sigma}\frac{d\sigma}{d\gamma_T} \propto
\frac{1}{\gamma_T^{\kappa} - 1} \simeq
\gamma_T^{-\kappa},\label{dgammaT}
\end{equation}
where for $a=1$ and $b=3$ one has $\kappa = 5$. These value seem
to be a reasonable first guess in the case of nucleon-nucleon
collisions ($b = 2m-1$, where $m$ denotes the number of spectator
quarks, whereas the parameter $a$ is given by the expected mean
value $\langle x\rangle = 1/3$). The exact form of Eq.
(\ref{dgammaT}) for $a=1$ and $b=3$ would be
\begin{equation}
\frac{1}{\sigma}\frac{d\sigma}{d\gamma_T} = 400
\frac{r^2}{\sqrt{\gamma^2_T - 1}}\left( \frac{1}{35} -
\frac{3r}{70} + \frac{r^2}{42} - \frac{r^3}{210}\right)
\end{equation}

with
\begin{equation}
r = \tan \phi_{02} = 2\gamma^2_T - 1 - 2 \gamma_T\sqrt{\gamma^2_T
- 1}.\label{eqb}
\end{equation}

\section{Fluctuations of relativistic temperature}

We close these remarks with the following observation. In Eq.
(\ref{AB4}) one encounters the scale factor $T\gamma_T$, not
$\gamma_T$ alone. Some time ago such a quantity was proposed as a
relativistic temperature \cite{RT},
\begin{equation}
T^{*}=\gamma_T T, \label{RT}
\end{equation}
(albeit in this case it would rather be a relativistic transverse
temperature, whatever that means). From the discussion following
Eq. (\ref{h2}) it follows that to get a proper Tsallis
distribution when fluctuating the scale parameter, one has to use
the full gamma function, Eq. (\ref{Gd}). Therefore, fluctuating
$T\gamma_T$ should bring the distribution of the transverse
momenta to the desired final Tsallis form. This, however, would
mean fluctuation of the relativistic temperature, $T^{*}$, but so
far its proper form is still under discussion \cite{BV}.
Nevertheless, keeping this reservation in mind let us elaborate on
such a possibility in more detail.

Assuming a beta distribution for fluctuations of $1/\gamma_T$,
\begin{equation}
b\left(\frac{1}{\gamma_T}\right) = \frac{1}{\Gamma(\kappa +
1)\Gamma(\alpha + 1)} \left(
\frac{1}{\gamma_T}\right)^{\kappa}\left( 1 -
\frac{1}{\gamma_T}\right)^{\alpha}, \label{beta}
\end{equation}
and a gamma distribution for fluctuations of $1/T$,
\begin{equation}
g\left(\frac{1}{T}\right) =
\frac{nT_0}{\Gamma(n)}\left(\frac{nT_0}{T}\right)^{n-1}\exp\left(
- \frac{nT_0}{T}\right), \label{gamma}
\end{equation}
one gets the joint distribution
\begin{equation}
g'\left(\frac{1}{T},\frac{1}{\gamma_T}\right) =
b\left(\frac{1}{\gamma_T}\right)\cdot g\left(\frac{1}{T}\right),
\label{joint}
\end{equation}
and can define the function
\begin{eqnarray}
g'\left(\frac{1}{T^{*}},\frac{1}{\gamma_T}\right) &=&
\frac{\left(nT_0\right)^n}{\Gamma(n)\Gamma(\kappa +
1)\Gamma(\alpha)}\cdot
\nonumber\\
&\cdot& \left(\frac{1}{T^{*}}\right)^n\left(\frac{1}{T} -
\frac{1}{T'}\right)^{\alpha}\exp\left( - \frac{nT_0}{T}\right).
\label{GTstarT}
\end{eqnarray}
In the case when the parameters of the two components of this
joint distribution are related in a certain way, for example if
$\alpha = n - 1 - \kappa$, fluctuations of the relativistic
temperature $T^{*}$ are again given by a gamma distribution (but
this time with a changed shape parameter):
\begin{eqnarray}
\!\!\!\!\!\!\!\!\!\! g'\left(\frac{1}{T^{*}}\right)\!\!\!
&=&\!\!\! \int_{1/T^{*}}^\infty
g'\left(\frac{1}{T^{*}},\frac{1}{T}\right)d\left(\frac{1}{T}\right) =\nonumber\\
&=&\!\!\! \frac{n - 1 - \kappa}{\Gamma(\kappa + 1)\Gamma(n +
1)}\left(\frac{nT_0}{T^{*}}\right)^{\kappa} \exp\left( -
\frac{nT_0}{T^{*}}\right) .\label{flucttstar}
\end{eqnarray}
This means that one could therefore consider the existence of
fluctuations of the relativistic temperature $1/T^{*}$, which are
again given by a gamma distribution with parameter $\kappa$
defining the size of these fluctuations. Relative fluctuations,
defined as
\begin{equation}
\omega(z) = \frac{Var\left(1/z\right)}{\langle 1/z\rangle^2},
\label{reffluct}
\end{equation}
satisfy the relation
\begin{equation}
\omega\left(T^{*}\right) = \frac{\omega(T) +
\omega\left(\gamma_T\right) + 2
\omega(T)\omega\left(\gamma_T\right)}{1 + 3 \omega(T)}
\label{omega}
\end{equation}
which connects fluctuations of $T$ and $\gamma_T$~~ \footnote{This
is another example of the generalized thermodynamic fluctuation
relations discussed previously in \cite{qFluct}.}. The
corresponding nonextensivity parameter is
\begin{equation}
q^{*} - 1 = \frac{1}{n^{*}} = \omega\left( T^{*}\right).
\label{qstar}
\end{equation}
Note that the parameter $n$, which defines fluctuations of
temperature in the decaying cluster, cf. Eq. (\ref{Gd}), is
proportional to heat capacity of the system under constant volume,
$n \propto C_V$. In the case considered here it is much greater
than $\kappa$, $n >> \kappa$. On the other hand, as shown in
\cite{Biro,Zakopane,CV}, the observed power index in the
corresponding Tsallis distribution is smaller than $n$. The above
consideration provides a simple and natural explanation of this
fact, the power index is defined by both $n$ and $\kappa$ (i.e. by
fluctuations of the relativistic temperature).

\section{Concluding remarks}

To summarize, the statistical cluster decay mechanism proposed in
\cite{AB} and discussed above, is yet another example of
supertatistics (which is not necessarily connected with a Tsallis
distribution). Fluctuations of the Lorentz factor given by the
distribution $g\left(\gamma_T\right)$ fully specify the slope
parameter of the transverse momentum distribution in the region of
large values of $p_T$. If they are given by a scale free
power-like distribution, as in Eq. (\ref{gamTfluct}), the
resultant distribution of $p_T$ is also a scale-free, power-like
one. Its behavior for small values of $p_T$ is dictated by the
fact that the Lorentz factor is defined only for $\gamma_T \ge 1$,
therefore there is always a natural cut-off in $g\left(
\gamma_T\right)$, eliminating $\gamma_T < 1$. As a result one gets
a distribution which is not a Tsallis distribution, remaining,
however, quite close to it numerically.

One can therefore invent a more general distribution, which is a
product of a beta distribution in $1/\gamma_T$ and a gamma
distribution for the parameter $1/T$. If the parameters defining
these distribution are related in some specific way, the resultant
distribution is again a gamma distribution with a modified shape.
In this case one gets the true Tsallis distribution for the
spectrum of the transverse momenta with the slope parameter
determined by $n^{*}$ from Eq. (\ref{qstar}). Note that in this
case it can be argued that it is determined by fluctuations of the
relativistic temperature $T^{*}= \gamma_T T$ (modulo reservations
concerning its proper form as mentioned before). Combining this
result with the connection between fluctuations of temperature and
specific heat, $\omega (T) \simeq 1/C_V$, one realizes that
fluctuations $\omega\left( \gamma_T\right)$ provide a natural
source of fluctuations resulting in the observed values of the
slope parameter $n^{*}$, which is smaller than $n \simeq C_V$. In
such a way the puzzle discussed in \cite{Biro,Zakopane,CV} could
find its natural solution.

We close by concluding that, as shown here, the realistic
superstatistics for high energy physics could have deviations from
the standard Gamma distribution and hence from Tsallis statistics,
although in an approximate sense Tsallis statistics is still
recovered. This provides yet another support for the idea
expressed in \cite{BBeck} that general superstatistics, based on
more general distributions than Gamma distribution, might be
relevant in high energy physics.\\

\vspace*{0.3cm} \noindent {\bf Acknowledgments}

This research  was supported in part by the National Science
Center (NCN) under contract Nr 2013/08/M/ST2/00598. We would like
to thank warmly Dr Nicholas Keely for reading this manuscript.

\end{document}